\documentclass[aps,amsmath,amssymb,amsfonts,preprint,nofootinbib]{revtex4}

\usepackage{graphicx}

\usepackage{tikz}
\usepackage{caption}
\usepackage{subcaption}
\usepackage{mathtools}

\usetikzlibrary{shapes.misc}
\usetikzlibrary{arrows.meta}
\tikzset{cross/.style={cross out, draw=black, minimum size=2*(#1-\pgflinewidth), inner sep=0pt, outer sep=0pt},
cross/.default={1pt}}
\tikzstyle{axisarrow} = [-{Latex[inset=0pt,length=5pt]}]

\usepackage[colorlinks=true]{hyperref} 
\hypersetup{
    bookmarks=true,         
    unicode=false,          
    pdftoolbar=true,        
    pdfmenubar=true,        
    pdffitwindow=false,     
    pdfstartview={FitH},    
    pdfsubject={},   
    pdfcreator={},   
    pdfproducer={}, 
    pdfkeywords={} {} {}, 
    pdfnewwindow=true,      
    colorlinks=true,       
    linkcolor=magenta, 
    citecolor=blue,        
    filecolor=magenta,      
    urlcolor=blue           
} 

\usepackage[export]{adjustbox}

\newcommand{\RR}{{\mathbb R}}
\newcommand{\ra}{\rightarrow}
\newcommand{\eps}{\epsilon}

\newcommand{\Tr}{{\rm Tr}}

\newcommand{\ga}{{\overset{\frown}{g}}}
\newcommand{\fa}{{\overset{\frown}{f}}}

\usepackage{amsmath}

\begin{document}

\title{Microscopic formulas for thermoelectric transport coefficients in lattice systems}
\author{Anton Kapustin}
\email{kapustin@theory.caltech.edu}
\author{Lev Spodyneiko}
\email{lionspo@caltech.edu}
\affiliation{California Institute of Technology, Pasadena, CA 91125, United States}


\begin{abstract}
A macroscopic description of thermoelectric phenomena involves several tensorial transport coefficients. Textbook microscopic Kubo formulas for them are plagued with ambiguities in the definitions of the current operators and the magnetization. We derive a version of these formulas for lattice systems which is  free from ambiguities but  contains additional terms compared to the textbook results. For symmetric components of  thermoelectric tensors, we identify a large class of lattice systems for which the additional terms vanish with a natural choice of the energy current. To eliminate ambiguities in the skew-symmetric components, one needs to interpret them as  relative quantities: only their differences for pairs of materials are well-defined. 
  
\end{abstract}

\maketitle

\section{Introduction}

Thermoelectric effects have many scientific and technological applications \cite{Goldsmid}. They can also serve as probes of novel materials. Thus it is important to develop a theoretical framework for computing thermoelectric coefficients in the most general setting, including strongly interacting materials without well-defined quasi-particles. 

Traditionally, the starting point for microscopic transport theory is provided by Kubo formulas. These formulas express transport coefficients in terms of correlators of volume-averaged current densities of conserved quantities. But although Kubo formulas go back to \cite{Kubo,Luttinger} and can be found in many textbooks and monographs \cite{Mahan,ZlaticMonnier}, there are a number of subtleties in their derivation. It is well appreciated by now that naive Kubo formulas for skew-symmetric parts of the transport tensors must be supplemented with additional terms involving magnetization and "energy magnetization" \cite{CHR}.  Such terms affect thermal Hall conductivity and the  skew-symmetric parts of thermoelectric coefficients. Since magnetizations are intrinsically ambiguous, it is not obvious how to evaluate such terms, see \cite{Niuetal} for a thorough discussion of magnetizations in general, \cite{XiaoNiu,XiaoRenXiong} for the semiclassical case, and \cite{GromovAbanov,ZhangGaoXiao} for geometric approaches to defining energy magnetization. Another rarely discussed issue is the ambiguity in the definition of the  energy density. One might expect that transport coefficients, being measurable quantities, are not affected by this ambiguity, but as far as we know this has been demonstrated only for the thermal conductivity and only for a special class of systems \cite{thermalgaugeinvariance}. 

The theory of transport coefficients developed in \cite{Luttinger,CHR,Niuetal} applies to continuum systems. It cannot be directly applied to lattice systems because it assumes certain scaling relations for electric and energy currents which do not hold on a lattice. (In fact, they do not hold for interacting continuum systems either, except after some spatial averaging \cite{CHR}). An even more basic issue is the lack of an accepted  definition of charge and energy current densities on a lattice. Many expositions of linearized transport theory (see e.g. \cite{Mahan,ZlaticMonnier}) derive only the expressions for the volume-averaged current densities. But in order to define transport coefficients one needs to separate currents into transport and magnetization contributions \cite{CHR,Niuetal}. Such a separation does not make sense for volume-averaged currents.


Since tight-binding models and other lattice Hamiltonians are ubiquitous in theoretical condensed matter physics, it is important to develop a formalism for describing currents of conserved quantities in such systems. In fact, such a formalism has been described by A. Kitaev many years ago \cite{kitaev}, but it is rarely applied to transport theory. In our recent work we used it to prove a Bloch theorem for energy currents \cite{energyBloch} and to derive Kubo-type formulas for the electric Hall conductivity and thermal Hall conductivity of general lattice systems  \cite{thermalHallpaper}. In this paper use the same approach to derive microscopic formulas for thermoelectric coefficients of general lattice systems.

The main results of the paper are as follows. Our formulas for the symmetric parts of conductivity and thermal conductivity tensors \cite{thermalHallpaper} are completely analogous to continuum formulas. Surprisingly, this not the case for the symmetric parts of thermoelectric tensors. In general, microscopic formulas for them contain local terms as well as the expected Kubo term. We show that these extra terms are in fact required to ensure that transport coefficients are unaffected by the ambiguities in the definition of the microscopic energy density. We also show that in special cases, such as systems of free particles or systems with only density-dependent interactions, the additional terms vanish with a natural definition of currents.

The skew-symmetric parts of all transport tensors except conductivity contain contributions from magnetizations. Since magnetizations are defined only up to additive constants, this leads to ambiguities. In the case of thermal conductivity, a way to resolve the ambiguities on a lattice  was described in \cite{thermalHallpaper} (building on the results of \cite{CHR,Niuetal}), and the same approach works for thermoelectric tensors. Namely, although skew-symmetric parts of these tensors are "contaminated" with edge effects, ambiguities cancel when one considers differences of transport tensors for two materials. We express this by saying that skew-symmetric tensors are relative transport coefficients. Microscopic formulas for relative transport coefficients take a more complicated form: they are integrals of differential 1-forms along a path in the space of parameters. These issues do not affect the skew-symmetric part of the conductivity tensor because one can, in principle, measure it in a torus geometry, where no boundaries are present. This is not possible to do even in principle for the skew-symmetric parts of other transport coefficients.

The content of the paper is as follows. In Section \ref{sec: hydro}, we explain how ambiguities in definition of transport and magnetization currents leads to a natural separation of transport coefficients into absolute and relative ones. In Section \ref{sec: themoelectric Kubo}, we derive  microscopic formulas for thermoelectric transport coefficients. We end with a discussion of possible generalizations of our results in Section \ref{sec: thermoel discussion}. In one of the appendices, we specialize our formulas to the case of non-interacting fermions and express thermoelectric coefficients in terms of zero-temperature 1-particle Green's functions in coordinate space.

The work was supported in part by the U.S.\ Department of Energy, Office of Science, Office of High Energy Physics, under Award Number DE-SC0011632. A.\ K.\ was also supported by the Simons Investigator Award.

\section{Relative and absolute transport coefficients}\label{sec: hydro}
The total current densities are usually divided into two parts:
\begin{align}\label{eq: tot into mag + tr}
\begin{split}
    {\bf j}^N_{\rm tot} &=     {\bf j}^N +     {\bf j}^N_{\rm mag},\\
      {\bf j}^E_{\rm tot} &=     {\bf j}^E +     {\bf j}^E_{\rm mag},
\end{split}
\end{align}
where magnetization currents are by definition divergence-free vector fields which do not contribute to net currents across any section of the system. Therefore they must have the form 
\begin{align}
    {\bf j}^N_{\rm mag} &=     \nabla \times    {\bf M}^N,\\
      {\bf j}^E_{\rm mag} &=      \nabla \times    {\bf M}^E,
\end{align}
where ${\bf M}^{N,E}$ are defined by these equations and are usually called magnetization density and energy magnetization density, respectively.  In the following we will use the same term magnetization for both magnetization and magnetization density which should not lead to a confusion.

The magnetization currents can be present even in an equilibrium state. The transport currents ${\bf j}^{N,E} $, on the other hand, can be present only in a non-equilibrium steady state created by slowly varying  gradients of external electric potential and temperature (for simplicity of presentation we assume the chemical potential to be constant). This follows from the Bloch theorem \cite{Watanabe,Bloch} and its energy analogue \cite{energyBloch}. This constrains the form of transport currents. 

Further constraints arise from gauge-invariance. It requires the transport  electric current ${\bf j}^N$ and the transport heat current\footnote{The correction terms in the heat current originate from $dN$ contributions to the heat $TdS = dE - \mu dN - \varphi dN$, where the last term represents the work done by electromagnetic field.} ${\bf j}^E-(\varphi+\mu) {\bf j}^N$ to be invariant under shifts of the electrical potential $\varphi$ by a constant. This follows from the equations defining the current operators~${\bf J}^{N,E}({\bf r})$
\begin{equation}\label{eq: hydro currents def}
i[ H, h({\bf r})]=-\nabla\cdot {\bf J}^E({\bf r}),\quad  i[ H, \rho({\bf r})]=-\nabla\cdot {\bf J}^N({\bf r}),
\end{equation}
and fact that under a constant gauge transformation $\varphi({\bf r})\mapsto\varphi({\bf r})+c$ the energy density also transforms as $h({\bf r})\mapsto h({\bf r})+c \rho({\bf r})$, where $\rho$ is the electric charge density.

Taking all these considerations into account, one finds that to leading order in the derivative expansion the transport electric current is given by 
\begin{equation}\label{eq: jn hydro}
{ j}^N_k=-\sigma_{km}\partial_m\varphi -\nu_{km} \partial_m T,
\end{equation}
where the conductivity tensor $\sigma_{km}$ and the thermoelectric tensor $\nu_{km}$ are functions of temperature only. For the energy current the expansion is 
\begin{equation}\label{eq: je hydro}
 { j}^E_k=(\varphi+\mu) { j}^N_k-\eta_{km}\partial_m\varphi -\kappa_{km} \partial_m T.
\end{equation}

A crucial point for this paper is that the separation of the current densities in (\ref{eq: tot into mag + tr}) is ambiguous. One can always remove a  curl of a vector field from ${\bf j}^{N,E}$ and add it to ${\bf j}^{N,E}_{\rm mag}$ without affecting the conservation equation  and the form of the transport equations (\ref{eq: jn hydro},\ref{eq: je hydro}).  While this should have no effect on physically observable quantities, it can affect the transport coefficients. Let us specialize to the 2d case and decompose all two-index tensors into symmetric and anti-symmetric parts: $\sigma_{km}=\sigma^S_{km}+\eps_{km}\sigma^A$ and similarly for the tensors $\nu,\eta,$ and $\kappa$. Taking into account the requirement of gauge-invariance, the allowed redefinitions of the transport currents have the form
\begin{align}\label{currenttransf}
   j^N_k &\mapsto j^N_k + \epsilon_{km}\partial_m  \left(\sigma_0(\varphi+\mu) +  f(T)\right), \\
    j^E_k &\mapsto j^E_k + \epsilon_{km}\partial_m\left(\frac12\sigma_0 (\varphi+\mu)^2+ f(T)(\varphi+\mu)+ g(T)\right).
\end{align}
Here $\sigma_0$ is a constant and $f(T), g(T)$ are arbitrary functions of $T$. Simultaneously magnetizations are redefined as follows:
\begin{align}
M^N &\mapsto M^N-\sigma_0(\varphi+\mu)-f(T),\\
M^E &\mapsto M^E-\frac12\sigma_0(\varphi+\mu)^2-f(T)(\varphi+\mu)-g(T).
\end{align}
After the redefinition transport coefficients change:
\begin{align}\label{eq: antisym part shifts}
\begin{split}
\sigma^A&\mapsto\sigma^A-\sigma_0,\\
\nu^A&\mapsto\nu^A-\frac{df(T)}{dT},\\
\eta^A&\mapsto\eta^A-f(T), \\ 
\kappa^A&\mapsto \kappa^A - \frac{dg(T)}{dT}.
\end{split}
\end{align}
Using such a redefinition we can always set $\sigma^A(T)$ to vanish at $T=0$ and make $\kappa^A$ and $\eta^A$ vanish identically for any homogeneous material. Instead of setting  $\eta^A(T)=0$, one can choose to set $\nu^A(T)=0$ and use the remaining freedom to set $\eta^A(0)=0$. Note also that $\dfrac{d{\sigma}^A}{dT}$ and  $\nu^A-\dfrac{d\eta^A}{dT}$ are invariant under such redefinitions.

So far we have ignored the vector potential, or equivalently gauge transformations which depend on the spatial coordinates. Allowing such gauge transformations changes the analysis as follows. Transport electric current ${\bf j}^N$ and transport heat current ${\bf j}^E-(\varphi+\mu) {\bf j}^N$ are now required to depend on $\varphi$ only through the electric field $E_k=-\partial_k\varphi-\frac{\partial A_k}{\partial t}$. The only difference this makes is that only transformations (\ref{currenttransf}) with $\sigma_0=0$ are allowed. As a result, the Hall conductivity $\sigma^A$ is now free from ambiguities.

There is a natural way to fix ambiguities in $M^N$ and $M^E$ and therefore also in $\nu^A,\eta^A$ and $\kappa^A$  \cite{CHR,thermalHallpaper}. If we consider a material with a boundary, the magnetizations as well as all transport coefficients can be set to zero outside. This removes all ambiguities from transport coefficients, but obscures the fact that some transport coefficients are defined relative to vacuum, while others do no depend on any choices and can be measured in the bulk. We will call them relative and absolute transport coefficients, respectively. According to the above analysis, all symmetric transport coefficients as well as $\sigma^A$ are absolute, while $\nu^A,\eta^A,$ and $\kappa^A$ are relative. The combination $\nu^A-\dfrac{d\eta^A}{dT}$ is also absolute.

This distinction has consequences for the microscopic formulas that can be derived for transport coefficients which are usually called Kubo formulas \cite{thermalHallpaper}. As we just explained, determination of relative transport coefficients require considering a system with boundaries. On the other hand, as we show in the paper,  derivatives of relative transport coefficients with respect to temperature or the parameters of the Hamiltonian  involves only correlation functions of a system without boundary.  The values of relative transport coefficients for any particular material can be found by integrating this differential over the parameters and/or temperature. The non-uniqueness in the choice of the base point of the integral reflects the ambiguity in the definition of the magnetization currents and can be fixed by choosing the base point to be a trivial insulator. The resulting microscopic  formula for a relative transport coefficient is manifestly independent of the choice of boundary conditions at the cost of depending on the correlation functions of a whole family of systems which interpolates between  the system of interest and a trivial insulator. On the other hand, microscopic formulas for absolute transport coefficients depend only on the correlation function of the system at a fixed temperature and values of all parameters.


As a consistency check, let us verify that physical bulk quantities depend only on absolute transport coefficients. For the time derivatives of charge and energy densities we get
\begin{align}\label{eq: div jn}
\frac{\partial\rho^N}{\partial t}=-\nabla\cdot {\bf j}^N &=-\sigma_{km}^S \partial_k {\bf E}_m-\frac{{d\sigma}_{km}} {dT} {\bf E}_m \partial_k T-\nu^S_{km}\partial_k\partial_m T-\frac{{d\nu}^S_{km}}{dT} \partial_k T\partial_m T,\\
\begin{split}\label{eq: div je}
\frac{\partial\rho^E}{\partial t}=-\nabla\cdot {\bf j}^E &=-(\varphi+\mu)\nabla\cdot {\bf j}^N+\sigma^S_{km}{\bf E}_k{\bf E}_m-\left(\nu_{km}+\frac{d{\eta}_{mk}}{dT}\right) {\bf E}_k \partial_m T-\eta^S_{km} \partial_k{\bf E}_m\\
&+\kappa^S_{km}\partial_k\partial_m T+\frac{d{\kappa}^S_{km}}{dT}\partial_k T\partial_m T,
\end{split}
\end{align}
where the external fields are assumed to be time-independent and thus  $\nabla\times {\bf E}=0$. As expected, these time derivatives are unaffected by the transformations (\ref{eq: antisym part shifts}).

The time-derivative of entropy density is usually written as follows \cite{LandauLifshits}:
\begin{equation}\label{eq:entropy prod}
\frac{\partial s}{\partial t}=\frac{1}{T}\sigma^S_{km} {\bf E}_k {\bf  E}_m+\frac{1}{T^2}\kappa^S_{km} \partial_k T\partial_m T-\frac{1}{T^2}(T\nu_{km}+ \eta_{mk}) {\bf E}_k \partial_m T-\nabla\cdot {\tilde {\bf j}}^S_k,
\end{equation}
where the entropy current density is
\begin{equation}\label{entropycurrent}
{\tilde {\bf j}}^S_k=\frac{1}{T}\eta_{km} {\bf E}_m-\frac{1}{T}\kappa^S_{km} \partial_m T .
\end{equation}
The r.h.s. of eq. (\ref{eq:entropy prod}) seems to depend on some relative transport coefficients. However, if one redefines the entropy current as follows:
\begin{equation}
{\bf j}^S_k=\frac{1}{T}\eta^S_{km}{\bf E}_m-\frac{1}{T}\kappa^S_{km}\partial_m T,
\end{equation}
then one can write
\begin{equation}
\frac{\partial s}{\partial t}=\frac{1}{T}\sigma^S_{km} {\bf E}_k {\bf  E}_m+\frac{1}{T^2}\kappa^S_{km} \partial_k T\partial_m T-\frac{1}{T^2}\left(T\nu^S_{km}+ \eta^S_{mk}+\eps_{km}T\left(\nu^A-\frac{d{\eta}^A}{dT}\right)\right) {\bf E}_k \partial_m T-\nabla\cdot {\bf j}^S_k.
\end{equation}
It is manifest now that both the entropy production rate and the entropy current depend only on the absolute transport coefficients.



Since only absolute transport coefficients enter the expressions for the divergences of currents, measuring net currents through closed curves (or surfaces, if we are discussing a 3d material) does not allow to determine relative transport coefficients. This applies even to infinite curves with a boundary at infinity, provided $\varphi$ and $T$ tend to fixed values at infinity. The latter condition must be imposed to eliminate the contribution of magnetization currents. For example, if we compute the electric current $I^N_x$ through a vertical line $x=0$, then the contribution of $\nu^A$ drops out because
\begin{equation}
\int_{-\infty}^\infty dy\,  \nu^A \partial_y T dy=0. 
\end{equation}

The above considerations apply to a homogeneous material whose transport coefficients are constants. If one considers a heterogeneous material, such as an interface between two homogeneous ones, then the expressions for net currents will involve differences between relative transport coefficients. For example, consider a sample such that $\nu^A$ interpolates between $\nu^A_1$ for $y\ll 0$ and $\nu^A_2$ for $y\gg 0$. Suppose that the temperature is a function of $y$ only which is equal to $T_b$ throughout the interface region and approaches $T_\infty$ at $y\ra\pm\infty$. Then the $\nu^A$-dependent contribution to the net electric current in the $x$ direction is
\begin{align}
 -\int_{-\infty}^\infty dy\, \nu^A \partial_y T  &  = (T_b-T_\infty) (\nu_2^A - \nu_1^A),
\end{align}
Similarly, the contribution of $\kappa^A$ to the net heat current is 
$(T_b-T_\infty) (\kappa_2^A - \kappa_1^A).$
By creating an electric potential $\varphi$ which is equal to $\varphi_b$ in the interface region and approaches $\varphi_\infty$ at $y\ra\pm\infty$, one can also measure $\sigma^A_2-\sigma^A_1$ and $\eta_2^A-\eta^A_1$.

\begin{figure}
    \centering
    \includegraphics{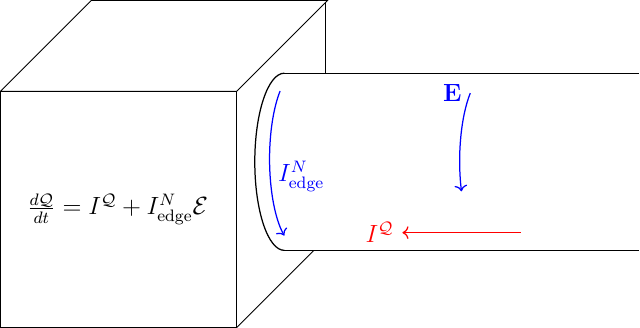}
    \caption{Insertion of a flux $\Phi$ into a cylinder creates electric field around it. If $\eta_{xy}\neq 0$, this electric field $\bf E$ drives a heat current $I^{\mathcal Q}$ along the axis of the cylinder. The jump of the magnetization gives rise to an edge current $I^N_{\rm edge}$. Work done on the edge current by the electric field contributes to the net heat $\mathcal Q$ transferred to the heat bath. 
    }
    \label{fig:flux}
\end{figure}

 In the case of the electric Hall conductivity one can do better  by utilizing a time-dependent vector potential rather than a scalar potential and working in a cylinder geometry or a  torus geometry. Then in principle one can determine $\sigma^A$ for a single material by measuring the net flow of electric charge across a section of a cylinder or a torus as one inserts a unit of magnetic flux through this section. This does not work for $\eta^A$ because the physical quantity that needs to be measured is the net amount of heat transferred to the heat bath as one inserts a unit of magnetic flux (see Fig. \ref{fig:flux}). Therefore heat transport will receive  a contribution from the work of the electromotive force $\mathcal E$ on the net electric edge currents $I^N_{\rm edge}$. The edge currents are proportional to the jump in the magnetization along the boundary and they make $\eta^A$ relative even in the  cylinder geometry. 

\section{Microscopic formulas for thermoelectric coefficients}\label{sec: themoelectric Kubo}
\subsection{Currents on a lattice}

We follow the conventions of \cite{thermalHallpaper}. We consider a lattice system with a Hamiltonian $H=\sum_{p\in\Lambda} H_p$, where $\Lambda\subset\RR^2$ is a not necessarily regular lattice.  The operators $H_p$ have a finite range, i.e. there exist $R$ such that $H_p$ acts trivially on site $q$ if $|p-q|>R$. The space of states at each site of the lattice is assumed to be finite-dimensional. The electric charge operator $Q$ has the form $Q=\sum_{p\in\Lambda} Q_p$, where $Q_p$ has integral eigenvalues (we set the electric charge of electron to be 1) and acts only on site $p$. This means that the $U(1)$ symmetry is on-site. In particular, $[Q_p,Q_q]=0$ for all $p,q\in\Lambda$. 

The electric current from site $q$ to site $p$ is defined as
\begin{equation}
J^N_{pq}=i[H_q,Q_p]-i[H_p,Q_q].
\end{equation}
The energy current from site $q$ to site $p$ is defined as 
\begin{equation}\label{energycurrent}
J^E_{pq}=-i[H_p,H_q].
\end{equation}
These currents enter the charge and energy conservation equations 
\begin{align}
    \frac{d Q_q}{dt} &= i\sum_{p\in \Lambda} [H_p,Q_q] = -\sum_{p\in \Lambda} J^N_{pq},\\
    \frac{d H_q}{dt} &= i\sum_{p\in \Lambda} [H_p,H_q] = -\sum_{p\in \Lambda} J^E_{pq}.
\end{align}

The net current from a subset $B\subset \Lambda$ to its complement $A=\Lambda \backslash B$ is given by
\begin{align}
    J^{N,E}(A,B) = \sum_{p\in A}\sum_{q\in B}  J^{N,E}_{pq}.
\end{align}
This observable measures the net current across the boundary of $A$ and $B$. In this paper we will need its mild generalization. For a given skew-symmetric function  $\eta(p,q): \Lambda \times \Lambda \rightarrow \mathbb R$ satisfying 
\begin{equation}
 \eta(p,q)+\eta(q,r) + \eta(r,p) = 0, \quad \forall p,q,r \in \Lambda,
\end{equation}
define 
\begin{align}\label{eq: J eta} 
    J^{N,E}(\eta) = \frac 1 2 \sum_{p,q\in \Lambda} \eta(p,q)J^{N,E}_{pq}.
\end{align}
Current $J^{N,E}(A,B)$ from $B$ to $A=\Lambda\backslash B$ corresponds to the case $\eta(p,q)= \chi_B(q)-\chi_B(p)$, where the function $\chi_B(p)$ is equal 1 on $B$ and 0 otherwise. For any function $\chi(p)$ we will denote by $\delta \chi(p,q) = \chi(q)-\chi(p)$ a function of two sites which can be thought of as a lattice analog of a gradient of the function $\chi$. For the function $\chi_B$, the operator $J^{N,E}(\delta \chi_B)$ measures the current through the boundary of region $B$.

The equilibrium expectation value of the currents satisfy
\begin{equation} \label{eq: div j = 0 lat}
\sum_{p\in \Lambda}\langle J^{N,E}_{pq} \rangle=0.
\end{equation}
This equation is a lattice analog of the continuum equation
\begin{align}
    \nabla \cdot \langle {\bf J}^{N,E}({\bf r})\rangle = 0.
\end{align}
In the continuum the general solution to this equation
\begin{align}\label{eq: cont magnetization}
    \langle J_k^{N,E}({\bf r})\rangle = -\epsilon_{kj}\partial_j M^{N,E}({\bf r})
\end{align}
defines the magnetization $M^N$ and energy magnetization $M^E$. Analogously, on a lattice the solution of equation (\ref{eq: div j = 0 lat}) is 
\begin{equation}\label{magnetization}
\langle J^{N,E}_{qr}\rangle=\sum_{p\in \Lambda} M^{N,E}_{pqr},
\end{equation}
where $M^{N,E}_{pqr}$ are skew-symmetric functions of the lattice points $p,q,r\in \Lambda$. These are lattice analogs of the magnetization and the energy magnetization. Physically, in continuum case $M^{N,E}(r)$ represent the circulating currents of the system in equilibrium. Similarly, $M^{N,E}_{pqr}$ physically can be thought as quantity which measures the circulating current around a triangle formed by $p,q,r$.

Unfortunately, $M^{N,E}_{pqr}$ is not unique: one can always redefine
\begin{equation}
M^{N,E}_{pqr}\mapsto M^{N,E}_{pqr}+\sum_{s\in \Lambda} N_{pqrs},
\end{equation}
where $N_{pqrs}$ is skewsymmetric function  of its subscripts which decays whenever any two of them are far apart.
This  corresponds to ambiguity in splitting of the circulating currents into contributions of magnetization from the different triangles and it is absent in continuum case. 

There is an additional ambiguity corresponding to existence of solutions to the equation $\sum_{r\in \Lambda}M^{N,E}_{pqr}=0$ which are not of the form $\sum_{s\in \Lambda} N_{pqrs}$. It corresponds to an ambiguity of addition of a constant to the magnetization in continuum case. A standard method to deal with the later is to consider a system with a boundary and fix the magnetization to be zero outside of the system. In this paper, we want to think about all transport coefficients as manifestly bulk quantities and avoid dealing with boundaries. While magnetization itself suffers from ambiguities and depends non-locally on the boundary conditions, the variation of magnetization with respect to parameters of the Hamiltonian is local. Indeed, consider the variation of the equation (\ref{magnetization}) with respect to a parameter $\lambda^\ell$ of the Hamiltonian
\begin{equation}\label{magnetizationderivative}
\frac{\partial}{\partial \lambda^\ell}\langle J^{N,E}_{pq}\rangle=\sum_{r\in\Lambda} \mu^{N,E}_{pqr,
\ell},
\end{equation}
    where $\mu^{N,E}_{pqr,\ell}=\dfrac{\partial M^{N,E}_{pqr}}{\partial \lambda^\ell}$ and is given by \cite{kitaev}

\begin{equation}\label{mue}
\mu^{N,E}_{pqr,\ell}=-\beta\langle\langle \frac{\partial H_p}{\partial \lambda^\ell}; J^{N,E}_{qr}\rangle\rangle-\beta\langle\langle \frac{\partial H_r}{\partial \lambda^\ell}; J^{N,E}_{pq}\rangle\rangle-\beta\langle\langle \frac{\partial H_q}{\partial \lambda^\ell}; J^{N,E}_{rp}\rangle\rangle,
\end{equation}
where $\langle\langle A;B\rangle\rangle$ denotes the Kubo canonical pairing \cite{Kubo}. Using the properties of the Kubo pairing (see Appendix A), 
one can easily verify the identity (\ref{magnetizationderivative}). In the following we will combine derivatives of magnetizations with respect to different parameters into 1-forms on the parameter space $\mu_{pqr}^{N,E}=\sum_{\ell} \mu^{N,E}_{pqr,\ell} d\lambda^{\ell}$.

\subsection{Equilibrium conditions and driving forces}
In the following sections we will follow Luttinger \cite{Luttinger} and study the behavior of the system coupled to external potentials
\begin{equation}
    H_p^{\psi,\varphi} =(1+\psi(p)) (H_p + \varphi(p) Q_p),
\end{equation}
where $\varphi(p)$ is external electric potential and $\psi(p)$ can be thought of as gravitational potential. The potentials are assumed to infinitesimally small slowly-varying functions of $p$ which vanish at infinity. After coupling to external potentials the system will eventually relax into a state with density matrix
\begin{equation}
    \rho \sim \exp\left( -\frac{H^{\psi,\varphi}-\mu_0 Q}{T_0}\right),
\end{equation}
where $T_0$ and $\mu_0$ are the temperature and local chemical potential of the system at infinity. On the other hand, on physical grounds we expect local observables supported in some small but macroscopic region around site $p$ to be described by a thermal density matrix
\begin{equation}
    \rho(p) \sim \left( -\frac{H-\mu(p) Q}{T(p)}\right),
\end{equation}
where the local temperature $T(p)$ and chemical potential $\mu(p)$ are slowly varying functions of $p$. The equilibrium conditions can be found to be \cite{Luttinger,CHR,Niuetal} 
\begin{align} \label{eq: elec eq}
    (1+\psi(p))(\mu(p)+\varphi(p)) &= \mu_0,\\ \label{eq: therm eq}
    (1+\psi(p)) T(p) &= T_0.
\end{align}
These relations together with the absence of transport currents in equilibrium can be used to derive Einstein relations between transport coefficients \cite{Luttinger}. The latter also leads to  transport currents being proportional to the gradients of the left hand sides of eqs. (\ref{eq: elec eq}) and (\ref{eq: therm eq}). The fact that the driving forces depend only on specific combinations of $\psi,\varphi,T, \mu $ will be used to relate the response to variations of the thermodynamic parameter $T$  to the  response to variations of the external field $\psi$.

\subsection{Nernst effect}
\begin{figure}[ht]
     \adjustbox{minipage=1.3em,valign=t}{\subcaption{}\label{fig: g kink fig 2d}}
      \begin{subfigure}[b]{0.45\textwidth}
    \centering
    \includegraphics[valign=t]{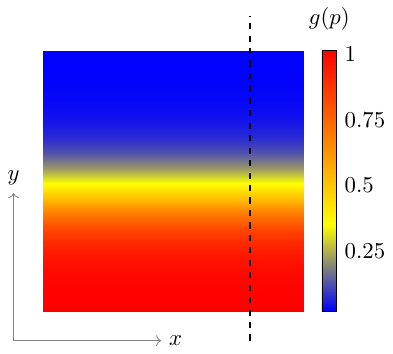}
\end{subfigure}
 \adjustbox{minipage=1.3em,valign=t}{\subcaption{}\label{fig: g hat fig 2d}}
  \begin{subfigure}[b]{0.45\textwidth}
    \centering
    \includegraphics[valign=t]{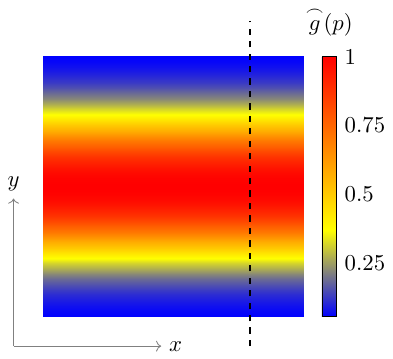}
\end{subfigure}
 \adjustbox{minipage=1.3em,valign=t}{\subcaption{}\label{fig: g kink fig}}
  \begin{subfigure}[b]{0.3\textwidth}
    \centering
    \includegraphics[valign=t]{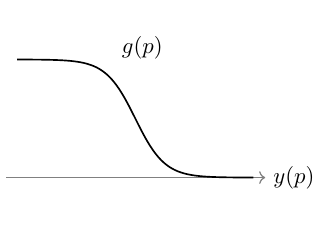}
\end{subfigure}
\hfill
 \adjustbox{minipage=1.3em,valign=t}{\subcaption{}\label{fig: g hat fig}}
\begin{subfigure}[b]{0.3\textwidth}
  \centering
  \includegraphics[valign=t]{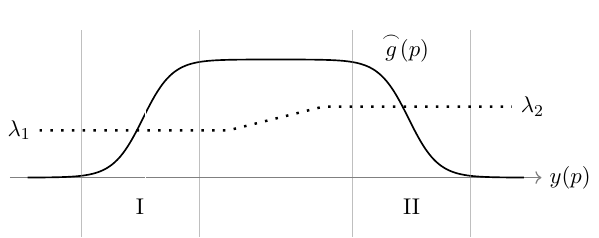}
\end{subfigure}
\hspace*{5cm} 
    \caption{ (a) Heat map of the function $g(p)$ corresponding to a temperature gradient in a horizontal strip. (b) Heat map of the function $\ga(p)$ corresponding to two horizontal strips with the opposite signs of the temperature gradient. (c) The function $g(p)$ restricted to the dashed line in (a). (d) The function $\ga(p)$ restricted to the dashed line in (b). Dotted line in (d) represents the dependence of some parameter $\lambda$ of the Hamiltonian on $y(p)$. }
    \label{fig:g}
\end{figure}

In order to find the thermoelectric coefficient coefficient $\nu_{xy}$ we deform the Hamiltonian density by
\begin{align}\label{eq: thermal pert}
    \Delta H_p = \epsilon e^{st} \ga(p) H_p,
\end{align}
where $\ga(p)$ is a hat-shaped function as in Fig. \ref{fig: g hat fig 2d}, $\epsilon$ is an  infinitesimal parameter, and $s$ is a small positive number which controls how fast the perturbation is turned on. We will consider the so-called fast regime \cite{Luttinger} in which the characteristic time $1/s$ is large but not large enough in order for the two slopes of the hat $\ga(p)$ to come into equilibrium.  

The change of the state of the system can be found as follows. The density matrix
\begin{align}
    \rho(t) = \rho_0 + \Delta \rho (t)
\end{align}
 satisfies the quantum Liouville equation
 \begin{equation}
    \frac{d \Delta \rho }{dt}= - i [\Delta H,\rho_0] + \dots,
 \end{equation}
where $\rho_0$ is the equilibrium density matrix at $t=-\infty$ and dots represent higher order terms in $\epsilon$. The solution to this equation is 
\begin{align}
    \Delta \rho (t=0) = - \rho_0 \int_0^\infty dt \int_0^\beta d\tau \Delta \dot H (-t-i\tau) ,
\end{align}
where the dot denotes the time derivative.
The change of the observable $A$ can be found to be
\begin{align}
    \Delta \langle A \rangle = \langle \Delta A\rangle -\beta \int_0^\infty dt   \langle \langle A;  \Delta \dot H (-t) \rangle \rangle,  
\end{align}
where we used the Kubo pairing notation (see Appendix \ref{app: Kubo}), $\beta=\frac 1 T$ is inverse temperature, and $\Delta A$ is the  variation of the operator arising from explicit dependence of $A$ on the Hamiltonian.
Using explicit form of the perturbation (\ref{eq: thermal pert}), energy conservation law and properties of the Kubo pairing we can rewrite this formula as
\begin{align}
    \Delta \langle A \rangle = \langle \Delta A\rangle +\epsilon\beta  \int_0^\infty dt e^{-st} \langle \langle A(t);  J^E(\delta \ga) \rangle \rangle,
\end{align}
where we neglected term proportional to small $s$. 

The change in the electric current across a vertical line $x=a$ is 
\begin{equation}
       \Delta \langle J^N(\delta f) \rangle = \langle \Delta J^N(\delta f) \rangle +\epsilon\beta  \int_0^\infty dt e^{-st} \langle \langle J^N(\delta f,t) ;  J^E(\delta \ga) \rangle \rangle,
\end{equation}
where $f(p)=\theta(a-x(p))$ is a step function. The explicit variation of the current is
\begin{align}
    \begin{split}
    &\langle \Delta J^N(\delta f) \rangle= \frac {i \epsilon} 2 \sum_{p,q\in \Lambda}\big\langle \ga(q)[H_q,Q_p] - \ga(p)[H_p,Q_q]\big\rangle (f(q)-f(p))\\&=\frac \epsilon 4 \sum_{p,q\in \Lambda} \langle J^N_{pq}\rangle(\ga(p)+\ga(q)) (f(q)-f(p))\\&\qquad + \frac {i\epsilon} 4 \sum_{p,q\in \Lambda} \langle [H_q,Q_p]+[H_p,Q_q](f(q)-f(p))(\ga(q)-\ga(p))\rangle,
    \end{split}
\end{align}
where in the last line we separated the result into two contribution formally skew-symmetric and symmetric in $f,\ga$. The second term depends only on difference of values of $\ga$ at different sites as expected for a   transport current. On the other hand, the first term depends on the value of $\ga$ and does not seem to be of the form expected for a transport current. As was explained in \cite{CHR} this contribution is related to magnetization currents. Indeed we can rewrite
\begin{align}\label{eq: j to mag}
    \frac \epsilon 4 \sum_{p,q\in \Lambda} \langle J^N_{pq}\rangle(\ga(p)+\ga(q)) (f(q)-f(p)) = \sum_{p,q,r\in \Lambda} M^{N}_{pqr} (f(q)-f(p))(\ga(q)-\ga(r)),
\end{align}
where we used skew-symmetry of $M_{pqr}^N$.

Written in this form, the response is proportional to the differences of $\ga$ at different points and therefore receives appreciable contributions only from regions $I$ and $II$ in Fig.~\ref{fig: g hat fig}. However, the transformation (\ref{eq: j to mag}) contains an important subtlety. The right-hand side hand side contains a magnetization contribution which is ambiguously defined while left-hand side is unambiguous. There is no contradiction because the ambiguity in the region $I$ will compensate the one in the region $II$. Moreover, in a homogeneous system the response will be zero, because these two regions compensate each other exactly. One way to deal with it is to introduce a boundary somewhere in between the two regions and enforce the magnetization $M$ to be 0 outside of the sample. This approach is used in \cite{CHR} in the continuum case. However, an explicit boundary introduces additional computational challenges and makes the bulk nature of the Nernst effect obscure. 

In this paper, we will use an alternative  approach proposed in  \cite{thermalHallpaper}. Instead of introducing a sharp boundary, we will make one parameter of the Hamiltonian $\lambda$ to have slightly different values in regions $I$ and $II$ (see Fig.~\ref{fig: g hat fig}). We can write the hat-shaped function $\ga$ as a difference of two functions $g_I$ and $g_{II}$, $\ga(p)=g_{II}(p)-g_I(p)$, each of which is a translate of the function $g(p)$ which depends only on $y(p)$ and is shown in Fig.~\ref{fig: g kink fig}. The functions $g_{I,II}$  are non-constant only in regions $I$ and $II$ respectively. Then the magnetization contribution can be rewritten as
\begin{equation}
     \epsilon (\lambda_{II}-\lambda_{I}) \mu^N_\lambda(\delta g\cup \delta f) + O((\lambda_{II}-\lambda_{I}) ^2),
\end{equation}
where $\mu^N_{pqr,\lambda} = \frac{\partial M^N_{pqr}}{\partial \lambda}$, and we introduced a notation
\begin{align}\label{eq: mag contraction}
    \mu^{N}_\lambda(\delta f_1 \cup \delta f_2) = \frac 1 6 \sum_{p,q,r\in\Lambda} \mu^N_{pqr,\lambda} (f_1(q)-f_2(p))(f_2(r)-f_2(q)).
\end{align}
Combining this with other contributions we find
\begin{align}\label{eq: nerst current}
\begin{split}
    \delta \langle J^N(\delta f)\rangle &\approx \epsilon (\lambda_{II}-\lambda_{I}) \Bigg\{\frac{\partial} {\partial \lambda}\Big[\beta  \int_0^\infty dt e^{-st} \langle \langle J^N(\delta f,t) ;  J^E(\delta g) \rangle \rangle +U(\delta f,\delta g)\Big]+ \mu^N_\lambda(\delta g\cup \delta f)\Bigg\},
\end{split}
\end{align}
where 
\begin{align}\label{eq: U def}
    U(\delta f,\delta g) = \frac {i} 4 \sum_{p,q\in \Lambda} \langle [H_q,Q_p]+[H_p,Q_q](f(q)-f(p))(g(q)-g(p))\rangle.
\end{align} 

The function $g$ as in Fig. \ref{fig: g kink fig} is not compactly supported and thus it takes an  infinite time for the system to equilibrate. Therefore, one can take the limit $s\rightarrow0$ while staying in the ``fast'' regime. 
Using the Einstein relation following from eqs. (\ref{eq: elec eq},\ref{eq: therm eq}), one finds that electric current after perturbation by gravitational potential $\epsilon g(p)$ is equal to the current generated by
\begin{align}
    T(p) &= \epsilon g(p) T_0,\\
    \varphi(p) &= \epsilon g(p) \mu_0.
\end{align}
From continuum phenomenological equation (\ref{eq: jn hydro}) one finds the current across $x=a$ line to be
\begin{equation}
    -\epsilon T_0 \int_{-\infty}^\infty \nu_{xy}\partial_y g dy  -\epsilon \mu_0 \int_{-\infty}^\infty \sigma_{xy} \partial_y g dy = \epsilon T_0 \int_{\lambda_I}^{\lambda_{II}} \frac{\partial\nu_{xy}}{\partial \lambda}d\lambda  + \epsilon \mu_0 \int_{\lambda_I}^{\lambda_{II}} \frac{\partial\sigma_{xy}}{\partial \lambda}d\lambda .
\end{equation}
Comparing it to (\ref{eq: nerst current}) we find
\begin{align}
\begin{split}
    d \nu_{xy} &=d\Big[\beta^2 \lim_{s\rightarrow0} \int_0^\infty dt e^{-st} \langle \langle J^N(\delta f,t) ;  J^{\mathcal Q}(\delta g) \rangle \rangle +\beta U(\delta f,\delta g) \Big]- \beta \mu^N(\delta f \cup\delta g ),
\end{split}
\end{align}
where we introduced the notation $J^{\mathcal Q}=J^E-\mu J^N$ for the heat current and we dropped the subscript 0 from $T_0$ and $\mu_0$ since this formula contains correlation functions of the unperturbed system in equilibrium. We combined differential with respect to parameter into the differential form $\mu^N(\delta f \cup\delta g ) = \mu^N_\lambda(\delta f \cup\delta g )d\lambda$ and the derivation can be straightforwardly extended to involve several parameters. The exterior derivative $d= \sum_\ell d\lambda \frac {\partial}{\partial \lambda^\ell}$ acts on the parameter space.

Since rescaling the temperature is equivalent to rescaling the Hamiltonian, we can extend this 1-form to the enlarged parameter space which includes $T$. Then we can define the difference of coefficients $\eta_{xy}$ for any two 2d materials, regardless of the temperature. Explicitly, let us define the rescaled Hamiltonian $H_{\lambda_0}=\lambda_0 H$, where we introduced an additional scaling parameter $\lambda_0$. Then
\begin{equation}
\left(T\frac{\partial}{\partial T}+\left.{\lambda_0}\frac{\partial}{\partial{\lambda_0}}\right|_{{\lambda_0}=1}\right)\nu_{xy}({\lambda_0},T)=0.
\end{equation}
Therefore we can define the $T$-component of the 1-form on the enlarged parameter space as follows:
\begin{equation} \label{eta Tderivative}
\frac{d\nu_{xy}}{dT}=\frac{\partial}{\partial T}\left[\beta^2  \int_0^\infty dt e^{-st} \langle \langle J^N(\delta f,t) ;  J^{\mathcal Q}(\delta g) \rangle \rangle +\beta U(\delta f,\delta g) \right]-\beta^2\tau^N(\delta f \cup \delta g),
\end{equation}
where $\tau^N(\delta f \cup \delta g)$ is given by eq. (\ref{eq: mag contraction}) with $\mu_{pqr}$ replaced with
\begin{equation}\label{eq: tau def} 
\tau^N_{pqr}=\beta\langle\langle H_p; J^N_{qr}\rangle\rangle+\beta\langle\langle H_r; J^N_{pq}\rangle\rangle+\beta\langle\langle H_q; J^N_{rp}\rangle\rangle,
\end{equation}
which is obtained from $\mu^N$ by replacing $dH_p$ with $-H_p$. 

\subsection{Ettingshausen effect}

One can derive a formula for the coefficient $\eta_{xy}$ in a similar way. In this section, we will display only the key steps, since all arguments are the same.

In order to find the coefficient we deform the Hamiltonian density by
\begin{align}
    \Delta H_p = \epsilon e^{st} \ga(p) Q_p,
\end{align}
where $\ga(p)$ is a hat-shaped function of $y(p)$ as in Fig.~\ref{fig: g hat fig 2d}.

The change in the energy current across a vertical line $x=a$ is 
\begin{equation}
       \Delta \langle J^E(\delta f) \rangle = \langle \Delta J^E(\delta f) \rangle +\epsilon\beta  \int_0^\infty dt e^{-st} \langle \langle J^E(\delta f,t) ;  J^N(\delta \ga) \rangle \rangle,
\end{equation}
where $f=\theta(a-x(p))$. The explicit variation of the current is
\begin{align}
    \begin{split}
    \langle \Delta J^E(\delta f) \rangle&= \frac {i \epsilon} 2 \sum_{p,q\in \Lambda}\big\langle \ga(p)[H_q,Q_p] - \ga(q)[H_p,Q_q]\big\rangle (f(q)-f(p))\\&\qquad=\frac \epsilon 4 \sum_{p,q\in \Lambda} \langle J^N_{pq}\rangle(\ga(p)+\ga(q)) (f(q)-f(p)) - U(\delta f, \delta \ga),
    \end{split}
\end{align}

The first term can be expressed in terms of magnetization as in (\ref{eq: j to mag}). We write $\ga(p)$ as a difference  $\ga(p)=g_{II}(p)-g_I(p)$, where $g_{I,II}$ are translates of a smeared step-function $g(p)$. Then we rewrite the response as a difference of conductivities of different materials:

\begin{align}\label{eq: etting current}
\begin{split}
    \delta \langle J^E(\delta f)\rangle &\approx \epsilon (\lambda_{II}-\lambda_{I}) \Bigg\{\frac{\partial} {\partial \lambda}\Big[\beta  \int_0^\infty dt e^{-st} \langle \langle J^E(\delta f,t) ;  J^N(\delta g) \rangle \rangle -U(\delta f,\delta g)\Big]+ \mu^N_\lambda(\delta g\cup \delta f)\Bigg\}.
\end{split}
\end{align}

On the other hand, from the continuum phenomenological equation (\ref{eq: je hydro}) one finds the current across the line $x=a$ to be
\begin{equation}
     -\epsilon  \int_{-\infty}^\infty \eta_{xy} \partial_y g dy  -\epsilon \mu \int_{-\infty}^\infty \sigma_{xy} \partial_y g dy =  \epsilon \int_{\lambda_I}^{\lambda_{II}} \frac{\partial\eta_{xy}}{\partial \lambda}d\lambda +\epsilon\mu \int_{\lambda_I}^{\lambda_{II}} \frac{\partial\sigma_{xy}}{\partial \lambda}d\lambda,
\end{equation}
where the second term originates from the contribution of $\varphi {\bf j}^N$ to ${\bf j}^E$. Comparing it to (\ref{eq: etting current}) we find
\begin{align}
\begin{split}
    d \eta_{xy} &=d\Big[\beta  \lim_{s\rightarrow0}\int_0^\infty dt e^{-st} \langle \langle J^{\mathcal Q}(\delta f,t) ;  J^N(\delta g) \rangle \rangle - U(\delta f,\delta g) \Big]- \beta \mu^N(\delta f \cup\delta g ).
\end{split}
\end{align}

This 1-form can be extended to include temperature as a parameter using the scaling relation
\begin{equation}
\left(T\frac{\partial}{\partial T}+\left.\lambda_0\frac{\partial}{\partial\lambda_0}\right|_{\lambda_0=1}\right) \frac {\eta_{xy}(\lambda_0,T)}{T}=0.
\end{equation}
Therefore we can define the $T$-component of the 1-form on the enlarged parameter space as follows:
\begin{equation}
\frac{d}{dT} \frac {\eta_{xy}}{T}=\frac{\partial}{\partial T}\left[\beta^2  \int_0^\infty dt e^{-st} \langle \langle J^{\mathcal Q}(\delta f,t) ;  J^N(\delta g) \rangle \rangle -\beta U(\delta f,\delta g) \right]-\beta^2\tau^N(\delta f \cup \delta g),
\end{equation}
where $\tau^N$ is given by (\ref{eq: tau def}).

\subsection{Symmetric parts of transport coefficients}

Note that the 1-forms $\mu^N(\delta f\cup\delta g)$ and $\tau^N(\delta f\cup\delta g)$ are formally skew-symmetric under the exchange of $f$ and $g$. To make use of this symmetry, we need to argue that $f$ can be replaced with a smeared step-function in the $x$ direction. This can be argued using  matching between microscopic theory and hydrodynamics. Namely, we expect that the microscopic linear response can be used to compute the properties of a Non-Equilibrium Steady State (NESS). Replacing $f$ with a smeared step-function changes the operators  $J^N(\delta f)$ and $J^E(\delta f)$ by $J^N(\delta \fa)$ and  $J^E(\delta \fa)$. The latter  operators can be equivalently written as $\frac{d Q(\fa)}{dt}$ and $\frac{d H(\fa)}{dt}$, where 
\begin{equation}
Q(\fa)=\sum_{p\in \Lambda} \fa(p) Q_p,\quad H(\fa)=\sum_{p\in \Lambda} \fa(p) H_p,
\end{equation}
and $\fa(p)$ is a hat-shaped function which depends only on $x(p)$. On the other hand, if the microscopic linear response is to reproduce the expected properties of a NESS, the expectation value of these observables must be zero. Thus we can replace $f$ with a smeared step-function in the $x$-direction without affecting $\nu_{xy}$ or $\eta_{xy}$. 

After this has been done, exchanging $f$ and $g$ is equivalent to exchanging $x$ and $y$. Thus $\mu^N(\delta f\cup\delta g)$ does not enter the microscopic formulas for the symmetrized thermoelectric coefficients. These formulas then can be integrated, giving 
\begin{align}  \nu^S_{xy} &=\frac{\beta^2}2  \lim_{s\rightarrow0} \int_0^\infty dt e^{-st} \left[\langle \langle J^N(\delta f,t) ;  J^{\mathcal Q}(\delta g) \rangle \rangle +\langle \langle  J^N(\delta g,t) ;  J^{\mathcal Q}(\delta f) \rangle \rangle \right]+\beta U(\delta f,\delta g),\\
\eta^S_{xy} &=\frac \beta 2  \lim_{s\rightarrow0}\int_0^\infty dt e^{-st} \left[\langle \langle J^{\mathcal Q}(\delta f,t) ;  J^N(\delta g) \rangle \rangle+\langle \langle J^{\mathcal Q}(\delta g,t) ;  J^N(\delta f) \rangle \rangle\right] - U(\delta f,\delta g).
\end{align}
As we show in Appendix \ref{App: invariance under Hamiltonian density redefinition}, the two terms on the right-hand side of these equation are not separately invariant under Hamiltonian density redefinition, but the full transport coefficients are invariant. The correction term $U(\delta f,\delta g)$ is zero for fermionic systems with only density-dependent interactions (see Appendix \ref{app: thermoelectric free fermions} for more details).

\subsection{Skew-symmetric parts of transport coefficients}

In a similar way one can find skew-symmetric parts of transport coefficients 
\begin{align}
        d \nu^A &=d\Big[\frac {\beta^2}2 \lim_{s\rightarrow0} \int_0^\infty dt e^{-st} \left( \langle \langle J^N(\delta f,t) ;  J^{\mathcal Q}(\delta g)\rangle \rangle - \langle \langle J^N(\delta g,t) ;  J^{\mathcal Q}(\delta f)\rangle \rangle\right) \Big]- \beta \mu^N(\delta f \cup\delta g ),\\
            d \eta^A &=d\Big[\frac \beta 2  \lim_{s\rightarrow0}\int_0^\infty dt e^{-st} \left(\langle \langle J^{\mathcal Q}(\delta f,t) ;  J^N(\delta g) \rangle \rangle-\langle \langle J^{\mathcal Q}(\delta g,t) ;  J^N(\delta f) \rangle \rangle \right) \Big]- \beta \mu^N(\delta f \cup\delta g ).
\end{align}
These formulas give only derivatives of transport coefficients with respect to parameters. Integration of these formulas over parameters or temperature gives the difference of relative transport coefficients at different values of parameters. It is natural to define relative transport coefficients of a trivial insulator to be zero. Determination of the relative transport coefficient in this case would correspond to integration over a path in the parameter space from a trivial insulator to the material of interest. 

There are many paths which one can use to deform a system into a trivial one. Consistency requires the integral of $d\nu^A$ or $d\eta^A$ to depend only on the endpoints of the path. This means that the 1-form $\mu^N(\delta f \cup\delta g )$ must be exact. This can be proved using the techniques of \cite{thermalHallpaper} where   $\mu^E(\delta f \cup\delta g )$ was  shown to be exact.

\section{Discussion}\label{sec: thermoel discussion}

In this paper we have derived microscopic formulas for ``transverse'' thermoelectric coefficients of general 2d lattice systems. It was convenient to decompose them into symmetric and anti-symmetric parts, since they have qualitatively different behavior: the former are absolute transport coefficients, while the latter are relative. Similar formulas for electric Hall conductivity and thermal Hall conductivity have already been derived in \cite{thermalHallpaper}. 

The usual Kubo formulas for transport coefficients require averaging the correlators of currents over the whole space. In contrast, our microscopic formulas involve net currents through two perpendicular lines. This is because a current on a 2d lattice is a function  of a pair of points, and the natural observable associated to it is localized on a line rather than at a point. Despite this, after the limit $s\ra 0$ has been taken, our formula computes the same quantity as the usual continuum Kubo formula. 

It is natural to ask whether longitudinal components of conductivity, thermal conductivity, and thermoelectric tensors of a 2d lattice system can be computed in a similar manner. This is easily achieved: one simply replaces two perpendicular lines with two lines making a nonzero angle $\theta$. It is easy to see determine from hydrodynamics which linear combination of components of the transport tensors describes the corresponding linear response. For example, if both currents involved are electric currents, one of the lines is given by $y=0$, and the other one is $y=x\cdot\tan \theta $, then the correlator
\begin{equation}
\beta \lim_{s\ra 0+}\int_0^\infty e^{-st} \langle\langle J^N(\delta f,t); J^N(\delta g)\rangle\rangle dt 
\end{equation}
measures $\sigma_{xy}-\frac{1}{\tan\theta}\sigma_{yy}$. Thus by changing the functions $f,g$ one can extract all four components of the conductivity tensor. The same is true about other transport coefficients.

In this paper we focused on the case of 2d materials, but the 3d case can be accommodated as well. One can simply replace a lattice in $\RR^2$ with a lattice in $\RR^2\times [0,L]$, impose periodic boundary conditions in the third direction, divide all formulas by $L$, and take the limit $L\ra\infty$. The functions $f,g$ remain independent of the third coordinate. It should not matter whether the limit $L\ra\infty$ is taken before or after the limit $s\ra 0$, since the problem is translationally invariant in the third direction.

In the 2d case, the quantity $\nu^A(T)$ (normalized relative to the vacuum) is dimensionless, and Bloch's theorem implies that for gapped systems $\nu^A(0)$ does not change under the variations of the Hamiltonian which do not close the gap. Thus if $\nu^A(0)$ were nonzero, it would represent a new topological invariant of gapped 2d phases of matter. However, one can show that on very general grounds  $\nu^A(0)$ vanishes for all gapped systems  \cite{Thirdlaw}. By Onsager reciprocity, the $T\ra 0$ limit of $\eta^A(T)/T$ also vanishes. Thus topological invariants of gapped 2d systems arise only from the Hall conductivity and the thermal Hall conductivity.

\appendix

\section{Kubo canonical pairing} \label{app: Kubo}

Kubo canonical pairing of two operators $A,B$ is defined as follows:
\begin{equation}
\langle\langle A;B\rangle\rangle=\frac{1}{\beta}\int_0^\beta\langle A(-i\tau)B\rangle d\tau-\langle A\rangle\langle B\rangle.
\end{equation}
Here $\langle\ldots\rangle$ denotes average over a Gibbs state at temperature $T=1/\beta$ (or more generally, over a state satisfying the Kubo-Martin-Schwinger condition), and $A(-i\tau)=e^{H\tau}A e^{-H\tau}$. Kubo paring determines static linear response: if the Hamiltonian is perturbed by $\lambda B$, where $\lambda$ is infinitesimal, then the change in the expectation value of $A$ due to the change in the equilibrium density matrix is
\begin{equation}
\Delta\langle A\rangle=-\beta\lambda\langle\langle A;B\rangle\rangle.
\end{equation}

Kubo pairing is symmetric, $\langle\langle A;B\rangle\rangle=\langle\langle B;A\rangle\rangle$, and satisfies
\begin{equation}
\beta\langle\langle i[H,A];B\rangle\rangle=\langle i[B,A]\rangle.
\end{equation}
In finite volume, one can write it in terms of the energy eigenstates as follows:
\begin{equation}
\langle\langle A;B\rangle\rangle=Z^{-1}\sum_{n,m} \langle n|\bar A|m\rangle\langle m|\bar B|n\rangle \frac{e^{-\beta E_m}-e^{-\beta E_n}}{\beta(E_n-E_m)},
\end{equation}
where $\bar A=A-\langle A\rangle,$ and $\bar B=B-\langle B\rangle.$

\section{Onsager reciprocity revisited}

Derivations of Onsager relations are  based on the analysis of hydrodynamic fluctuations, so it might seem that they  should put constraints only on those  transport coefficients which enter the  hydrodynamic equations of motion. On closer inspection, one finds \cite{Casimir} that the derivation involves net currents which measure the rate of change of conserved quantities in a finite volume and thus  require understanding boundary contributions. As a result, Onsager reciprocity constrains both absolute transport coefficients and relative transport coefficients defined relative to the vacuum. Equivalently, it imposes conditions on the derivatives of relative transport coefficients with respect to parameters. To illustrate how this works, let us discuss the constraints imposed by Onsager reciprocity on relative transport coefficients of time-reversal-invariant 2d  systems. For the skew-symmetric thermal conductivity $\kappa^A$, time-reversal-invariance implies
\begin{equation}
\frac{\partial}{\partial\lambda}\kappa^A=0,
\end{equation}
where $\lambda$ is a parameter of the Hamiltonian. Thus $\kappa^A$ can be a function of temperature only. Further, if we treat $T$ as a parameter, then scaling analysis gives
\begin{equation}
\frac{\partial}{\partial T} \frac{\kappa^A}{T}=0.
\end{equation}
Hence $\kappa^A(T)=a T$, where $a$ does not depend on parameters. The parameter $a$ has no physical significance, but it is natural to set it to zero, so that the vacuum has zero thermal Hall conductivity. Thus we reach the standard conclusion that for a system with time-reversal invariance $\kappa^A=0$.

The case of thermoelectric coefficients is slightly different. Usually one says that Onsager reciprocity requires $\nu_{km}=T^{-1}\eta_{mk}$, which implies $\nu^A+T^{-1}\eta^A=0$ \cite{LandauLifshits}.
Since both $\nu^A$ and $\eta^A$ are relative transport coefficients, one should interpret this as
\begin{equation}
\frac{\partial}{\partial\lambda} \left(\nu^A+T^{-1}\eta^A\right)=0.
\end{equation}
Hence $\nu^A+T^{-1}\eta^A$ can depend only on the temperature. If we treat temperature as a parameter, then the scaling analysis gives
\begin{equation}
\frac{\partial}{\partial T}\left(\nu^A+T^{-1}\eta^A\right)=0.
\end{equation}
Hence $\nu^A+T^{-1}\eta^A=a$, where $a$ is a constant which is independent of any parameters or temperature and has no physical significance. One can choose it to be zero. Then $\nu^A=-T^{-1}\eta^A$. So for a time-reversal-invariant 2d system there is only one independent skew-symmetric thermoelectric transport coefficient, namely $\nu^A$.


\section{Invariance under Hamiltonian density redefinition}\label{App: invariance under Hamiltonian density redefinition}
For a given Hamiltonian, there are many ways to define a  Hamiltonian density. A typical example of this is the  ambiguity in splitting an interaction term between two sites $p$ and $q$ into $H_p$ and/or $H_q$. In this appendix, we will show that our microscopic formulas for physically observable transport coefficients are independent of the choice of the Hamiltonian density, even though individual terms in the microscopic formulas are not invariant. For some systems this can be used to simplify the microscopic formulas.

\subsection{Invariance of the electric current}
Consider the following change of the Hamiltonian density 
\begin{align}\label{eq: hamiltonian gauge trans}
    H_p \rightarrow H_p + \sum_{r \in \Lambda} A_{rp},
\end{align}
where $A_{rp}$ is skew-symmetric in $r,p$. We want the final Hamiltonian to be $U(1)$-invariant. Therefore, we have to impose
\begin{equation}\label{eq: A condition}
    [Q,\sum_{r\in \Lambda} A_{rp}]=0.
\end{equation}
For a general choice of $A_{pq}$ a stronger condition
\begin{equation}\label{eq: A condition v2}
    [Q,A_{pq}]=0,
\end{equation}
will not hold. However, one can always redefine $A_{pq}$ (by subtracting the $U(1)$-non-invariant part) in such a way that ({\ref{eq: A condition v2}}) holds without affecting $H_p$. In the following we will assume this was done and (\ref{eq: A condition v2}) is true.

Under the transformation (\ref{eq: hamiltonian gauge trans}) the electric current changes as
\begin{align}\label{eq: el cur 2chain shift}
  J^N_{pq}\rightarrow   J^N_{pq}+  i\sum_{r\in \Lambda} \left([A_{rq},Q_p]-[A_{rp},Q_q]\right).
\end{align}
Even though the current density changes, the net current through any section is invariant. Indeed,
\begin{equation}
    J^N(\delta f) \rightarrow J^N(\delta f) +  \frac i  2\sum_{p,q,r\in \Lambda} \left([A_{rq},Q_p]-[A_{rp},Q_q]\right)(f(q)-f(p)),
\end{equation}
and the last term is zero since
\begin{multline}
    \sum_{p,q,r\in \Lambda} \left([A_{rq},Q_p]-[A_{rp},Q_q]\right)(f(q)-f(p))  \\= \sum_{p,q,r\in \Lambda} \left([A_{rq},Q_p]+[A_{pr},Q_q]+[A_{qp},Q_r]\right) (f(q)-f(p)) \\= \frac 1 3\sum_{p,q,r\in \Lambda} \left([A_{rq},Q_p]+[A_{pr},Q_q]+[A_{qp},Q_r]\right) (f(q)-f(p)+f(p)-f(r)+f(r)-f(q))=0,
\end{multline}
where we have used (\ref{eq: A condition v2}) and the  skew-symmetry of $[A_{rq},Q_p]+[A_{pr},Q_q]+[A_{qp},Q_r]$. 
\subsection{Covariance of the energy current}

Let us now consider the effect of the redefinition of the Hamiltonian density on the energy current. Imposing an  energy analog of (\ref{eq: A condition}) or (\ref{eq: A condition v2}) 
\begin{equation}\label{eq: A condition for H}
    [H,\sum_{r\in \Lambda} A_{rp}] \stackrel{?}{=}0, \qquad \textrm{or}\qquad [H,A_{pq}] \stackrel{?}{=}0,
\end{equation}
is far too restrictive, since it would only allow changes of the Hamiltoniain density by conserved quantities. For example,  the difference between putting the interaction term between the two sites $p$ and $q$  either into $H_p$ or into $H_{q}$ corresponds to $A_{pq}$ equal to the interaction term.  Obviously, interaction terms  are not integrals of motion in general. Because of this we will not impose either of the equations in (\ref{eq: A condition for H}).

Under the redefinition of the Hamiltonian density (\ref{eq: hamiltonian gauge trans}) the energy current changes as 
\begin{align}
    J^E_{pq}\rightarrow   J^E_{pq}+ i\sum_{r\in \Lambda} \left([A_{rq},H_p]+[H_q,A_{rp}]\right),
\end{align}
while the net current transforms as
\begin{align}\label{eq: energy cur  shift}
    J^E(\delta f)\rightarrow   J^E(\delta f)+ \frac i 2\sum_{p,q,r\in \Lambda} \left([A_{rq},H_p]+[H_q,A_{rp}]\right)(f(q)-f(p)).
\end{align}
The last term can be rewritten as
\begin{multline*}
     \frac i 2\sum_{p,q,r\in \Lambda} \left([A_{rq},H_p]+[H_q,A_{rp}]\right)(f(q)-f(p)) \\= \frac i 2 \sum_{p,q,r\in \Lambda} \left([A_{rq},H_p]+[A_{pr},H_q]+[A_{qp},H_r]\right)(f(q)-f(p)) - \frac i 2 \sum_{p,q,r\in \Lambda} [A_{qp},H_r](f(q)-f(p)) \\= \frac i 6\sum_{p,q,r\in \Lambda} \left([A_{rq},H_p]+[A_{pr},H_q]+[A_{qp},H_r]\right)(f(q)-f(p)+f(p)-f(r)+f(r)-f(q))  \\- \frac i 2 \sum_{p,q\in \Lambda}[H,A_{pq}](f(q)-f(p)) = -\dot A (\delta f ),
\end{multline*}
where we have defined 
\begin{align}
     A (\delta f ) = \frac 1 2 \sum_{p,q\in \Lambda}A_{pq}(f(q)-f(p)).
\end{align}

We find that the net energy current transforms as follows  under a redefinition of the Hamiltonian density:
\begin{align}\label{eq: net energy cur  shift}
    J^E(\delta f)\rightarrow   J^E(\delta f)-\dot A (\delta f ).
\end{align}
But this should be expected since a redefinition of the energy density changes  how we define the energy of  sub-regions and therefore should affect the net energy current. Indeed, one can see that (\ref{eq: energy cur  shift}) is exactly the transformation needed in order to satisfy the energy conservation law
\begin{align}
    \dot H_p = -\sum_{q\in \Lambda} J^E_{pq} \quad \rightarrow\quad \dot H_p +\sum_{q\in \Lambda} \dot A_{qp} = -\sum_{q\in \Lambda} J^E_{pq} +\sum_{q\in \Lambda} \dot A_{qp}
\end{align}
for the new energy density $H_p +\sum_{q\in\Lambda} A_{qp}$. By summing this transformation law over $p$ weighted by a function $f(p)$ with a compact support we find that
\begin{align}
    \dot H(f) = -J^E(\delta f)\quad \rightarrow\quad \dot H_p +\dot A(\delta f) = -J^E(\delta f) +\dot A(\delta f),
\end{align}
which reproduces (\ref{eq: net energy cur  shift}). Here we used an identity  
\begin{align}
    \sum_{p,q\in \Lambda} A_{pq} f(q) = \frac 1 2  \sum_{p,q\in \Lambda} A_{pq} ( f(q) -f(p)) = A(\delta f )
\end{align}
which is true for any $f$ with a compact support.

From the above discussion, one can see that energy current is not invariant but covariant under energy density redefinitions. If we choose $f(p)$ to be 1 when $p$ is in some compact set $B$ and zero otherwise,  the physical meaning of  (\ref{eq: net energy cur  shift}) is very clear. It corresponds to ambiguities in the energy currents due to interaction terms along the boundary of $B$. Depending on how we distribute the interaction terms among $H_p$ we can change the energy stored in the region $B$ as well as energy current  through its boundary.

\subsection{Invariance of the microscopic formulas for  thermoelectic coefficients}

In this section we will show that the coefficients $\nu_{xy}$ and $\eta_{xy}$ are invariant under a redefinition of the Hamiltonian density. We will start with skew-symmetric coefficients
\begin{align}
    d\nu^A&=\frac 1 2 d\Big( \nu^{\rm Kubo}(\delta f, \delta g)- \nu^{\rm Kubo}(\delta g, \delta f)\Big)-\beta^2 \mu^N(\delta f \cup \delta g),\\
    d\eta^A&=\frac 1 2 d\Big(\eta^{\rm Kubo}(\delta f, \delta g)- \eta^{\rm Kubo}(\delta g, \delta f)\Big)-\beta \mu^N(\delta f \cup \delta g) .
\end{align}
Here we defined the Kubo parts as
\begin{align}\label{eq: Kubo part nu}
    \nu^{\rm Kubo}(\delta f, \delta g) &= \beta^2 \lim_{s\rightarrow0} \int_0^\infty dt e^{-st} \langle \langle J^N(\delta f,t) ;  J^{\mathcal Q}(\delta g) \rangle \rangle,\\\label{eq: Kubo part eta}
    \eta^{\rm Kubo}(\delta f, \delta g)&=\beta  \lim_{s\rightarrow0}\int_0^\infty dt e^{-st} \langle \langle J^{\mathcal Q}(\delta f,t) ;  J^N(\delta g) \rangle \rangle.
\end{align}

Under Hamiltonian density redefinition the Kubo parts transform as 
\begin{align}
\begin{split}
    \nu^{\rm Kubo}(\delta f, \delta g) &\rightarrow\nu^{\rm Kubo}(\delta f, \delta g) -\beta^2 \lim_{s\rightarrow0} \int_0^\infty dt e^{-st} \langle \langle J^N(\delta f,t) ; \dot A(\delta g) \rangle \rangle\\&=\nu^{\rm Kubo}(\delta f, \delta g) -\beta^2\langle\langle J^N(\delta f);  A(\delta g)\rangle\rangle, 
    \end{split} \\
    \begin{split}
    \eta^{\rm Kubo}(\delta f, \delta g) &\rightarrow\eta^{\rm Kubo}(\delta f, \delta g) -\beta \lim_{s\rightarrow0} \int_0^\infty dt e^{-st} \langle \langle  \dot A(\delta f,t); J^N(\delta g) ; \rangle \rangle\\&=\nu^{\rm Kubo}(\delta f, \delta g) +\beta\langle\langle A(\delta f);  J^N(\delta g)\rangle\rangle, 
    \end{split} 
\end{align}
where we used properties of the Kubo pairing.

Before finding the variation of the magnetization term it is useful to rewrite it slightly:
\begin{align}\label{eq: half magnetization}
\begin{split}
    \mu^N(\delta f \cup \delta g) =\frac 1 2 \sum_{p,q\in\Lambda} \left[\frac 1 3 \sum_{r\in\Lambda} \mu_{pqr}(g_p+g_q+g_r) - \frac 1 2 \sum_{r\in\Lambda} \mu_{pqr} (g(p)+g(q))\right] (f(q)-f(p))\\= \frac 1 2 \sum_{p,q\in\Lambda} \left[\frac 1 3 \sum_{r\in\Lambda} \mu_{pqr}(g_p+g_q+g_r) - \frac 1 2 d\langle J^N_{pq}\rangle (g(p)+g(q))\right] (f(q)-f(p)).
\end{split}
\end{align}
Note that one cannot expand the square brackets, since the two resulting sums over $p,q$ will not converge separately.

Let us find the variation of $ \dfrac 1 2 \langle J^N_{pq}\rangle (g(p)+g(q))$ under a Hamiltonian density redefinition. It reads
\begin{align}\label{eq: mag gauge transf}
\begin{split}
\frac 1 2 \langle J^N_{pq}\rangle (g(p)+g(q))&\rightarrow  \frac 1 2 \langle J^N_{pq}\rangle (g(p)+g(q))\\&+ \frac i 2 \sum_{r\in \Lambda}\langle [ A_{rq}, Q_p] -[ A_{rp}, Q_q] \rangle (g(p)+g(q))
\end{split}
\end{align}
The last term can be rewritten as follows:
\begin{align}\label{eq: comm to kubo}
\begin{split}
     \frac i 2 \sum_{r\in \Lambda} \langle [ A_{rq}, Q_p] &-[ A_{rp}, Q_q] \rangle (g(p)+g(q)) \\&= \frac \beta 2\sum_{r\in \Lambda}\langle\langle g(p)\dot Q_p ;  A_{rq} \rangle \rangle+\frac \beta 2\sum_{r\in \Lambda}\langle\langle \dot Q_p ;g(q)  A_{rq}  \rangle \rangle - (p\leftrightarrow q),
\end{split}
\end{align} 
where we used the properties of the Kubo pairing. The first term in this expression can be rewritten as
\begin{multline}\label{eq: cap to ev 1}
\sum_{r\in \Lambda} \langle\langle g(p)\dot Q_p; A_{rq} \rangle\rangle- (p\leftrightarrow q) = -\frac 1 2 \sum_{s,r\in\Lambda}\langle\langle J_{sp}^N(g(s)+g(p));  A_{rq}  \rangle \rangle - (p\leftrightarrow q) \\= -\frac  1 2 \sum_{s,r\in\Lambda}\langle\langle  J^N_{sp}(g(s)+g(p)) + J^N_{ps} (g(s)-g(p)); A_{rq}  \rangle \rangle - (p\leftrightarrow q) =\langle\langle J^N(\delta g); A_{qp} \rangle \rangle\\- \frac 1 2  \sum_{r,s\in\Lambda}  \Bigg[\langle\langle J^N_{rp}(g(r)+g(p)); A_{sq} \rangle \rangle+\langle\langle J^N_{ps}(g(s)-g(p)); A_{rq} \rangle \rangle+\text{2 perms}\Bigg] ,
\end{multline}
where "$\text{2 perms}$" means the two cyclic permutations in $p,q,r$. Note that the term in square brackets is skew-symmetric in $p,q,r$. The second term can be rewritten as
\begin{multline}\label{eq: cap to ev 2}
\sum_{r\in \Lambda}\langle\langle \dot Q_p; g(q)  A_{rq} \rangle \rangle- (p\leftrightarrow q) =- \sum_{s,r\in \Lambda}\langle\langle  J^N_{sp}; g(q)   A_{rq} \rangle \rangle- (p\leftrightarrow q)\\=
-\frac 1 2 \sum_{s,r\in \Lambda}\langle\langle  J^N_{sp};  A_{rq}(g(q)+g(r)) +A_{qr}(g(r)-g(q)) \rangle \rangle- (p\leftrightarrow q)=\langle\langle  J^N_{pq}; A (\delta g) \rangle \rangle\\-\frac 1 2 \sum_{s,r\in \Lambda}\Big[\langle\langle  J^N_{sp}; A_{rq}(g(q)+g(r))\rangle \rangle+\langle\langle  J^N_{rp}; A_{qr}(g(r)-g(q))\rangle \rangle+\text{2 perms}\Big].
\end{multline}
Note that term in square brackets is skew-symmetric in $p,q,r$

By combining equations (\ref{eq: half magnetization}-\ref{eq: cap to ev 2}) we find that the magnetization contribution changes under a redefinition of the Hamiltonian density as follows:
\begin{align}
\begin{split}
    \mu^N(\delta f \cup \delta g)  \rightarrow \mu^N(\delta f \cup \delta g) -\frac \beta 2 d \langle\langle  J^N(\delta f); A (\delta g) \rangle \rangle&+\frac \beta 2 d\langle\langle  J^N(\delta g); A (\delta f) \rangle \rangle \\&+\frac 1 2 \sum_{p,q,r\in\Lambda} C_{pqr}(f(q)-f(p)),
\end{split}
\end{align}
where $C_{pqr}$ is a skew-symmetric function of $p,q,r$ which is combination of skew-symmetric parts (and their derivatives) in the right-hand sides of Eqs. (\ref{eq: half magnetization}-\ref{eq: cap to ev 2}). Due to its skew-symmetry we find that
\begin{equation}
    \frac 1 2 \sum_{p,q,r\in\Lambda} C_{pqr}(f(q)-f(p)) = \frac 1 6 \sum_{p,q,r\in\Lambda} C_{pqr}(f(q)-f(p)+f(p)-f(s)+f(s)-f(q))=0.
\end{equation}

We see that the variation of the magnetization exactly compensates the variation of the Kubo parts. Thus the  skew-symmetric parts of the thermoelectric tensors are invariant under a redefinition of the Hamiltonian density.

Now let us consider the symmetric parts
\begin{align}
    \nu^S_{xy} &=\frac 1 2 \Big( \nu^{\rm Kubo}(\delta f, \delta g)+ \nu^{\rm Kubo}(\delta g, \delta f)\Big)+\beta U(\delta f , \delta g),\\
    \eta^S_{xy}&=\frac 1 2 \Big(\eta^{\rm Kubo}(\delta f, \delta g)+ \eta^{\rm Kubo}(\delta g, \delta f)\Big)- U(\delta f , \delta g).
\end{align}
The variation of Kubo parts were already determined before, so we focus on the transformation of $U$. Under (\ref{eq: hamiltonian gauge trans}) it transforms as follows:
\begin{align}
   U(\delta f , \delta g) \rightarrow U(\delta f , \delta g)+ \frac {i} 4 \sum_{p,q\in \Lambda} \langle [\partial A_q,Q_p]+[\partial A_p,Q_q]\rangle(f(q)-f(p))(g(q)-g(p)).
\end{align}
We can rewrite this equation by noticing that
\begin{align}
    \frac {i} 2  \langle [\partial A_q,Q_p]+[\partial A_p,Q_q]\rangle(g(q)-g(p)) = -\frac \beta 2\langle\langle g(p)\dot Q_p ; \partial A_q \rangle \rangle+\frac \beta 2\langle\langle \dot Q_p ;\partial g(q)  A_q  \rangle \rangle - (p\leftrightarrow q).
\end{align}
Then using eqs. (\ref{eq: cap to ev 1}, \ref{eq: cap to ev 2}) we find 
\begin{equation}
      U(\delta f , \delta g) \rightarrow U(\delta f , \delta g)  +\frac \beta 2  \langle\langle  J^N(\delta f); A (\delta g) \rangle \rangle+\frac \beta 2 \langle\langle  J^N(\delta g); A (\delta f) \rangle \rangle
\end{equation}
We see that the variation of this term cancels the varitions of the Kubo parts.

One can do the same checks for the thermal Hall conductivity and verify that the microscopic formula derived in \cite{thermalHallpaper} is in invariant under a redefinition of the Hamiltonian density. To linear order in $A_{pq}$ all the manipulations are almost the same except for the  replacement $Q_p\rightarrow H_p$ and $J^N \rightarrow J^E$.

\section{Thermoelectric coefficients for free fermions}\label{app: thermoelectric free fermions}
\subsection{Definitions and correlation functions}
In this appendix we will specialize our microscopic formulas for coefficients $\nu$ and $\eta$ to free fermionic systems. The Hamiltonian is taken to be
\begin{equation}\label{eq: free Hamiltonioan}
    H=\sum_{p,q\in \Lambda} a^\dagger_p h(p,q) a_q,
\end{equation}
where an infinite matrix $h(p,q)$ is Hermitian $h(p,q)^*=h(q,p)$, and $a^\dagger_p, a_p$ are fermionic creation-annihilation operators satisfying the standard anti-commutation relations
\begin{equation}
    a_p a^\dagger_q + a^\dagger_q a_p = \delta_{p,q}, \qquad   a_p a_q + a_q a_p =   a_p^\dagger a^\dagger_q + a^\dagger_q a_p^\dagger=0.
\end{equation}

We define the Hamiltonian density on site $p$ to be
\begin{equation}
H_p=\frac12\sum_{m\in \Lambda} \left(a^\dagger_p h(p,m)a_m+a^\dagger_m h(m,p) a_p\right).
\end{equation}
The charge operator on site $p$ is defined as 
\begin{equation}
    Q_p = a^\dagger_pa_p.
\end{equation}
The electric current can be found from the  conservation equation:
\begin{equation}
    J_{pq}^N = i (a^\dagger_q h(q,p) a_p - a^\dagger_p h(p,q) a_q).
\end{equation}
The net current through a section defined by $\delta f (p,q)=f(q)-f(p)$  is 
$$
J(\delta f)=-i a^\dagger [h,f] a,
$$
where a bounded function $f\in\ell^2(\Lambda)$ is understood as an operator acting on the one-particle Hilbert space $\ell^2(\Lambda)$ by multiplication. Summation over sites is implicit.

The energy current operator is
\begin{multline}
J^E_{pq}=\frac{-i}{4}\sum_{m\in\Lambda}\left(a^\dagger_ph(p,q)h(q,m) a_m-a^\dagger_q h(q,p)h(p,m) a_m\right.\\
\left. -a^\dagger_m h(m,q)h(q,p) a_q+a^\dagger_m h(m,p)h(p,q) a_q\right.\\
\left. +a^\dagger_ph(p,m)h(m,q)a_q-a^\dagger_q h(q,m)h(m,p)a_p\right).
\end{multline}
The net energy current is
$$
J^E(\delta f)=-\frac{i}{2} a^\dagger [h^2,f] a.
$$
The state of the system at a temperature $T=1/\beta$ is defined via Wick's theorem and Gibbs distribution
\begin{eqnarray}
\langle a_p(t) a^\dagger_q \rangle &= &\left\langle p\left\vert \frac{e^{-ih t}}{1+e^{-\beta h}}\right\vert q\right\rangle, \\
\langle a_p(t)^\dagger a_q\rangle &= & \left\langle q\left\vert \frac{e^{ih t}}{1+e^{\beta h}}\right\vert p\right\rangle,
\end{eqnarray}
where $a_p(t)$ are operators in  the Heisenberg picture.

Using these formulas we find
$$
\langle J^N(\delta f,t) J^N(\delta g)\rangle=- {\rm Tr}\left( [h,f] \frac{e^{-iht}}{1+e^{-\beta h}} [h,g] \frac{e^{iht}}{1+e^{\beta h}}\right),
$$
where the trace is over the 1-particle Hilbert space $\ell^2(\Lambda)$, and the functions $f:\Lambda\ra \RR$ and $g:\Lambda\ra \RR$ are operators on this Hilbert space. The operators $[h,f]$ and $[h,g]$ have support on a vertical strip and  a horizontal strip, respectively. 

Switching to the energy basis, substituting $t\ra t-i\tau$, and integrating from $0$ to $\beta$ over $\tau$  we find
$$
\langle\langle J^N(\delta f,t); J^N(\delta g)\rangle\rangle=\frac{-1}{\beta} \sum_{n,m} \langle n\vert [h,f]\vert m\rangle \langle m\vert [h,g]\vert n\rangle e^{i(\varepsilon_n-\varepsilon_m) t} \frac{e^{\beta \varepsilon_n}-e^{\beta \varepsilon_m}}{(1+e^{\beta \varepsilon_n})(1+e^{\beta \varepsilon_m})(\varepsilon_n-\varepsilon_m)},
$$
where $\varepsilon_n$ are 1-particle Hamiltonian energy eigenvalues.

Multiplying this by $e^{-st}$ and integrating over $t$, we arrive at
$$
\sigma_{xy}=i\lim_{s\rightarrow 0}\sum_{n,m} \frac{\langle n\vert [h,f]\vert m\rangle \langle m\vert [h,g]\vert n\rangle}{\varepsilon_n-\varepsilon_m+is} \frac{\mathfrak f(\varepsilon_n)-\mathfrak f(\varepsilon_m)}{\varepsilon_n-\varepsilon_m} ,
$$
where $\mathfrak f(\varepsilon)=\frac{1}{1+e^{\beta( \varepsilon)}}$ is the Fermi-Dirac distribution. We absorb the chemical potential into a shift of the Hamiltonian.

The above expressions assume a discrete energy spectrum and thus can only be used for finite-volume systems. To get an expression applicable to infinite-volume systems,
let us rewrite it in terms of the one-particle Green's functions $G_\pm(z)=1/(z-h\pm i0)$. Some of the useful formulas are
$$
\langle a^\dagger A a\rangle=-\frac 1 {2\pi i}\int_{-\infty}^\infty dz \, \mathfrak f(z) \Tr  \Big(\big[G_+-G_-\big] A\Big),
$$
\begin{align}
    \begin{split}
      -\beta\langle\langle a^\dagger A a; a^\dagger B a\rangle\rangle=-\frac 1 {2\pi i}\int_{-\infty}^\infty dz\, \mathfrak f(z)\Tr \Big(\big[G_+-G_-\big] A G_+ B + G_-  A\big[G_+-G_-\big] B\Big) \\=  -\frac 1 {2\pi i}\int_{-\infty}^\infty dz\, \mathfrak f(z)\Tr \Big(G_+ A G_+ B - G_-  A G_- B\Big) ,
    \end{split}
\end{align}
where we have dropped $z$ for $G_\pm(z)$. Here $A$ and $B$ are operators acting on the one-particle Hilbert space, and in the second formula we assumed in addition that their average is zero: $\langle a^\dagger A a\rangle=\langle a^\dagger B a\rangle =0$. Note also that
$$
hG_{\pm}=G_{\pm}h=zG_{\pm}-1,\quad [G_\pm,A]=G_\pm[h,A]G_\pm.
$$
Using this notation the formula for the electric conductivity takes the form
\begin{multline}\label{free case sigma}
\sigma_{xy}=-\frac 1 {2\pi} \int_{-\infty}^{\infty} dz \,\mathfrak f(z) \Tr\big\{ [h, f] G_+^2 [h, g] (G_+-G_-)-[h, f](G_+-G_-)[h, g]  G_-^2 \big\},
\end{multline}
where the integration is over the real axis in the $z$-plane.

\subsection{Magnetization term}
The magnetization differential for an arbitrary deformation $dh$ of the 1-particle Hamiltonian is given by 
\begin{multline}\label{variation of magnetization free system}
\mu^N(\delta f\cup\delta g)=\frac 1 {4\pi}\int_{-\infty}^\infty dz \,\mathfrak f(z)\Tr\Big( G_+dhG_+\Big\{ [h,f]G_+[h,g]-[h,g]G_+[h,f]\Big\}\Big)-(G_+ \rightarrow G_-).
\end{multline}

For temperature variations this expression can be simplified to
\begin{multline}
\tau^N(\delta f\cup\delta g)=\frac 1 {4\pi}\int_{-\infty}^\infty dz \Tr\Big(  \mathfrak f(z)(G_+-G_-)[h^2,f]G_-^2[h, g]-\mathfrak f(z)(G_+-G_-)[h^2, g]G_+^2[h,f]\\+\mathfrak f'(z)(G_+-G_-)h[h, g]G_+[h,f]+\mathfrak f'(z)(G_+-G_-)h[h, g]G_-[h,f]\Big) - (f\leftrightarrow g).
\end{multline}
These expressions are needed only for the evaluation of skew-symmetric parts of the thermoelectric coefficients.

\subsection{$U$-term}
Let us study the term (\ref{eq: U def}) for free fermionic system. In this case the relevant many-body operators become
\begin{equation}\label{eq: U free commutator}
\left([Q_p,H_q]+[Q_q,H_p]\right) =a^\dagger[h,\delta_p \delta_q] a,
\end{equation}
where $\delta_p$ is a Kronecker delta function equal 1 on site $p$ and 0 on all other sites. A product of two delta functions enforces $q=p$ in the summation over $p$ and $q$. Since $U$ also involves a factor of $(g(p)-g(q))(f(p)-f(q))$, $U(\delta f,\delta g)$ vanishes for systems of free fermions.

More generally, one can consider a system of fermions with only density-dependent interactions. Namely, suppose we allow the following interaction term in the Hamitonian (\ref{eq: free Hamiltonioan}):
\begin{equation}
    H^{\rm int}= \sum_{p_1,\dots,p_n \in \Lambda} V(p_1,\dots,p_n) Q_{p_1} \dots Q_{p_n},
\end{equation}
where $V(p_1,\dots,p_n)$ is a function of $n$ sites which describes the potential energy of many-body interaction and decays rapidly when the points $p_1,\ldots, p_n$ are far from each other.
One can see that this term will leave eq. (\ref{eq: U free commutator}) unaffected since $Q_p$ commute with each other. We conclude that for fermionic system with only density-dependent interactions there is no correction originating from $U$  to the symmetric thermoelectric coefficients provided $H_p$ is chosen in the manner explained above.

\subsection{Skew-symmetric part}
Consider the variation of the  Kubo parts (\ref{eq: Kubo part nu},\ref{eq: Kubo part eta}) of  the skew-symmetric thermoelectic coefficients under a rescaling of the Hamiltonian: $dh=h\,d\lambda_0$. We get
\begin{multline}
d\nu^A_{\rm Kubo}=\beta d\eta^A_{\rm Kubo}(\delta f\cup\delta g)=\frac {d\lambda_0} {4\pi}\int_{-\infty}^\infty dz \Tr\Big(  \mathfrak f(z)(G_+-G_-)[h^2,f]G_-^2[h, g]\\-\mathfrak f(z)(G_+-G_-)[h^2, g]G_+^2[h,f]+\mathfrak f'(z)(G_+-G_-)h[h, g]G_+[h,f]+\mathfrak f'(z)(G_+-G_-)h[h, g]G_-[h,f]\\-2\mathfrak f'(z)h^2(G_+-G_-)[h,f]G_+^2[h, g]+2\mathfrak f'(z)h^2(G_+-G_-)[h, g]G_-^2[h,f]\Big) - (f\leftrightarrow g).
\end{multline}

Summing up this contributions with the  magnetization contribution gives
\begin{multline}
\frac{d\nu^A}{dT}=\frac{d}{dT}\left( \frac{\eta^A}{T}\right)=\frac 1 {2\pi T^2}\int_{-\infty}^\infty dz\mathfrak f'(z)z^2 \Tr\Big( (G_+-G_-)[h,f]G_+^2[h, g]\\-(G_+-G_-)[h, g]G_-^2[h,f] - (f\leftrightarrow g)
\Big).
\end{multline}

Integrating over the temperature  and using the formula
\begin{align}
 \int_{T}^\infty \frac {dT} {T^2} \mathfrak f'(z) = \frac  {c_1(\mathfrak f(z))-\log 2} {z^2},
\end{align}
where 
\begin{align}
    c_1(x) = \int_0^x dt \log\left(\frac {1 - t} t\right) = - x \log x - (1-x) \log(1-x),
\end{align}
gives 
\begin{align}
\begin{split}
\nu^A=\frac{\eta^A}T =\frac 1 {2\pi}\int_{-\infty}^\infty dz\, c_1 \big(\mathfrak f(z)\big) \Tr\Big( (G_+&-G_-)[h,f]G_+^2[h, g]\\&-(G_+-G_-)[h, g]G_-^2[h,f]\Big) - (f\leftrightarrow g).
\end{split}
\end{align}
Here we normalized the thermoelectric coefficients to be 0 in the infinite-temperature state. Note that since in the limit $T\ra 0$ the Fermi-Dirac distribution $\mathfrak f(z)$ becomes a step-function, and since $c_1(0)=c_1(1)=0$, both $\nu^A(T)$ and $\eta^A(T)/T$ vanish at $T=0$ regardless of the choice of the Hamiltonian $h$. 

\subsection{Symmetric part}
Symmetric parts of transverse thermoelectric coefficients are 
\begin{align*}
    \nu^S_{xy} &= -\frac {\beta} {8\pi}  \int_{-\infty}^{\infty} dz \,\mathfrak f(z) \Tr\big\{ [h, f] G_+^2 [h^2, g] (G_+-G_-)-[h, f](G_+-G_-)[h^2, g]  G_-^2 \big\}+(f\leftrightarrow g),\\    
    \eta^S_{xy} &= -\frac 1 {8\pi} \int_{-\infty}^{\infty} dz \,\mathfrak f(z) \Tr\big\{ [h^2, f] G_+^2 [h, g] (G_+-G_-)-[h^2, f](G_+-G_-)[h, g]  G_-^2 \big\}+(f\leftrightarrow g).
\end{align*}
As explained in the body of the paper,  longitudinal parts are given by the same formulas with a more general choice of the functions $f,g$.

\bibliographystyle{apsrev4-1}
\bibliography{bib}

\end{document}